\documentclass[10pt,aps,pra,superscriptaddress,amsmath,amssymb,twocolumn,amsfonts,floatfix, longbibliography]{revtex4-2}
\usepackage{graphicx}
\usepackage{epstopdf}
\usepackage[T1]{fontenc}
\usepackage[latin9]{inputenc}
\usepackage{amsbsy}
\usepackage{gensymb}
\usepackage[dvipsnames]{xcolor}
\definecolor{deepfuchsia}{rgb}{0.76, 0.33, 0.76}
\definecolor{electricpurple}{rgb}{0.75, 0.0, 1.0}

\usepackage{graphicx}
\usepackage{epstopdf}
\usepackage[T1]{fontenc}
\usepackage[latin9]{inputenc}
\usepackage{amsbsy}
\usepackage{gensymb}
\usepackage[T1]{fontenc}
\usepackage[latin9]{inputenc}
\usepackage{amsmath}
\usepackage{amssymb}
\usepackage{bbm}
\usepackage{braket}
\usepackage{xcolor}
\allowdisplaybreaks
\usepackage{graphicx}
\usepackage[colorlinks=true]{hyperref}
\usepackage{orcidlink}
\usepackage[normalem]{ulem}
\newcommand{\equref}[1]{Eq.~(\ref{#1})}

\newcommand{\figref}[1]{Fig.~\ref{#1}}

\newcommand{\tableref}[1]{Table~\ref{#1}}

\renewcommand{\approx}{\simeq}

\newcommand{\beq}{\begin{equation}}
\newcommand{\eeq}{\end{equation}}
\newcommand{\bea}{\begin{eqnarray}}
\newcommand{\eea}{\end{eqnarray}}

\usepackage{mathtools}
\hypersetup{
    unicode=false,          
    pdftoolbar=true,        
    pdfmenubar=true,        
    pdffitwindow=false,     
    pdfstartview={FitH},    
    pdftitle={Unconventional superconductivity in a non-centrosymmetric ${\alpha}$-Mn alloy NbTaOs$_2$},    
    pdfauthor={R.K. Kushwaha, R. P. Singh},     
    pdfsubject={},   
    pdfcreator={},   
    pdfproducer={}, 
    pdfkeywords={} {} {}, 
    pdfnewwindow=true,      
    colorlinks=true,       
    linkcolor=blue, 
    citecolor=blue,        
    filecolor=magenta,      
    urlcolor=blue           
}
\begin{document}
\title{\textrm{Unconventional superconductivity in a non-centrosymmetric ${\alpha}$-Mn alloy NbTaOs$_2$}}
\author{{R.~K.~Kushwaha}\,\orcidlink{0009-0005-3457-3653}}
\affiliation{Department of Physics, Indian Institute of Science Education and Research Bhopal, Bhopal, 462066, India}
\author{{Arushi}\,\orcidlink{0000-0003-4400-5260}}
\affiliation{Department of Physics, Indian Institute of Science Education and Research Bhopal, Bhopal, 462066, India}
\author{S.~Jangid}
\affiliation{Department of Physics, Indian Institute of Science Education and Research Bhopal, Bhopal, 462066, India}
\author{P. K.~Meena}
\affiliation{Department of Physics, Indian Institute of Science Education and Research Bhopal, Bhopal, 462066, India}
\author{R.~Stewart}
\affiliation{ISIS Facility, STFC Rutherford Appleton Laboratory, Harwell Science and Innovation Campus, Oxfordshire, OX11 0QX, UK}
\author{{A. D. Hillier}\,\orcidlink{0000-0002-2391-8581}}
\affiliation{ISIS Facility, STFC Rutherford Appleton Laboratory, Harwell Science and Innovation Campus, Oxfordshire, OX11 0QX, UK}
\author{{R.~P.~Singh}\,\orcidlink{0000-0003-2548-231X}}
\email[]{rpsingh@iiserb.ac.in} 
\affiliation{Department of Physics, Indian Institute of Science Education and Research Bhopal, Bhopal, 462066, India}
\begin{abstract}
\begin{flushleft}
\end{flushleft}

Non-centrosymmetric superconductors have emerged as a fascinating avenue for exploring unconventional superconductivity. Their broken inversion and time-reversal symmetries make them prime candidates for realizing the intrinsic superconducting diode effect (SDE). In this work, we synthesize the ternary non-centrosymmetric $\alpha$-Mn alloy NbTaOs$_{2}$ and conduct a comprehensive investigation of its superconducting properties through resistivity, magnetization, specific heat and muon spin rotation/relaxation ($\mu$SR) techniques. Our transverse field-$\mu$SR and specific heat results provide evidence of a moderately coupled, fully-gaped superconducting state. Zero field-$\mu$SR measurements reveal a subtle increase in the relaxation rate below the transition temperature, suggesting time reversal symmetry breaking in the superconducting ground state of NbTaOs$_{2}$.
\end{abstract}
\maketitle
\section{Introduction}
The superconducting order parameter, defined by complex gap functions, reveals the macroscopic quantum state, with its symmetry crucial for superconductor classification. While conventional superconductors break only gauge symmetry U(1), unconventional superconductors can disrupt additional symmetries, including time-reversal, rotational, and inversion symmetries~\cite{RevModPhys.63.239}. Non-centrosymmetric superconductors (NCSs), lacking inversion symmetry, represent a distinct class of unconventional superconductors~\cite{CePt3Si,Smidman}. The resulting antisymmetric spin-orbit coupling (ASOC) in these materials enables mixing of the spin-singlet and spin-triplet pairing states~\cite{ASOC,Smidman}. This mixed pairing leads to unique phenomena in NCSs, such as nodes/nodal lines and multigap behavior~\cite{nl1,nl2,nl3}, an upper critical field exceeding the Pauli limit~\cite{highHc2}, and non-trivial topological states~\cite{topo1,topo2,HfRhGe}. 

Recent studies predict that NCSs exhibit a range of unique effects, including piezosuperconductivity~\cite{PhysRevB.105.134514}, chiral superconductivity~\cite{SrPtAs}, field-reversed vortices~\cite{PhysRevB.102.184516}, nonlinear responses~\cite{NLOR, tkachov2018probing}, and a superconducting analog of the Chandrasekhar-Kendall state~\cite{CKstate}. Notably, NCSs emerge as ideal materials in which to realize the intrinsic superconducting diode effect (SDE)~\cite{nadeem2023superconducting,PhysRevLett.128.037001}. Their inherently broken inversion and time-reversal symmetries provide a natural and simplified pathway to achieve SDE, in contrast to conventional approaches that rely on complex heterostructures and externally applied magnetic fields~\cite{narita2022field,intrisicSDE}.

While NCSs exhibiting time-reversal symmetry breaking (TRSB) are theoretically intriguing, they remain experimentally rare, highlighting the need for further discovery and characterization. TRSB in the superconducting state~\cite{TRSBreview} has been observed across various centrosymmetric (CS) and non-centrosymmetric (NC) crystal structures~\cite{Kagome,FeSePRL,Sr2RuO4PRl,luke1998time,cuprate,CaSb2,HfRhGe,Re6ZrTRSB,Re6Hf,Re6Ti,La7Ir3,La7Rh3,La7Ni3,UTe2,UPt3Luke}. Notably, rhenium-based NCSs with the $\alpha$-Mn structure reported a frequent occurrence of TRSB with fully gapped superconductivity~\cite{Re6ZrTRSB,Re6Hf,Re6Ti}. The critical percentage of rhenium in these compounds, along with the complex interplay of the crystal structure and ASOC, is believed to be critical for TRSB. This is further evidenced by the lack of TRSB in non-rhenium-based superconductors~\cite{NbOs,NbOs2,TaOs}. However, most of the work on NCS is focused on binary compounds, and TRSB is not universal in $\alpha$-Mn structured superconductors~\cite{Re3W,Re3Ta,Mg10Ir19B16,NbOs,NbOs2,TaOs}. This suggests that other factors are at play in determining TRSB presence or absence in NCSs, highlighting the need to discover and characterize new noncentrosymmetric superconductors.

We have synthesized a new ternary compound NbTaOs$_{2}$, which crystallizes in the non-centrosymmetric $\alpha$-Mn structure and is known for its intrinsic disorder. Due to the mixing of 4$d$ niobium and 5$d$ tantalum atoms on the atomic sites, we expect enhanced ASOC and further disorder. This will help to understand the complex interplay of crystal structure, ASOC, and the effect of disorder on the superconducting ground state. Furthermore, this study may provide insight into the role of the critical percentage of a specific element in determining the presence and absence of TRSB in the superconducting ground state~\cite{RePRL}.

 In this paper, we investigate the superconducting properties of NbTaOs$_{2}$ using electrical resistivity, magnetization, specific heat, and muon spin rotation/relaxation ($\mu$SR) measurements. Our findings confirm that NbTaOs$_{2}$ exhibits bulk type II superconductivity, with a superconducting transition temperature of 2.4(2)~K. Both the specific heat and transverse field $\mu$SR measurements confirm an isotropic superconducting gap. Zero-field $\mu$SR measurements show an increased relaxation rate below the superconducting transition temperature, suggesting a broken time-reversal symmetry in the superconducting ground state of NbTaOs$_{2}$. This makes NbTaOs$_{2}$ the first non-Re-based $\alpha$-Mn superconducting compound to exhibit time-reversal symmetry breaking.

\begin{figure*} [t]
\includegraphics[width=2.05\columnwidth, origin=b]{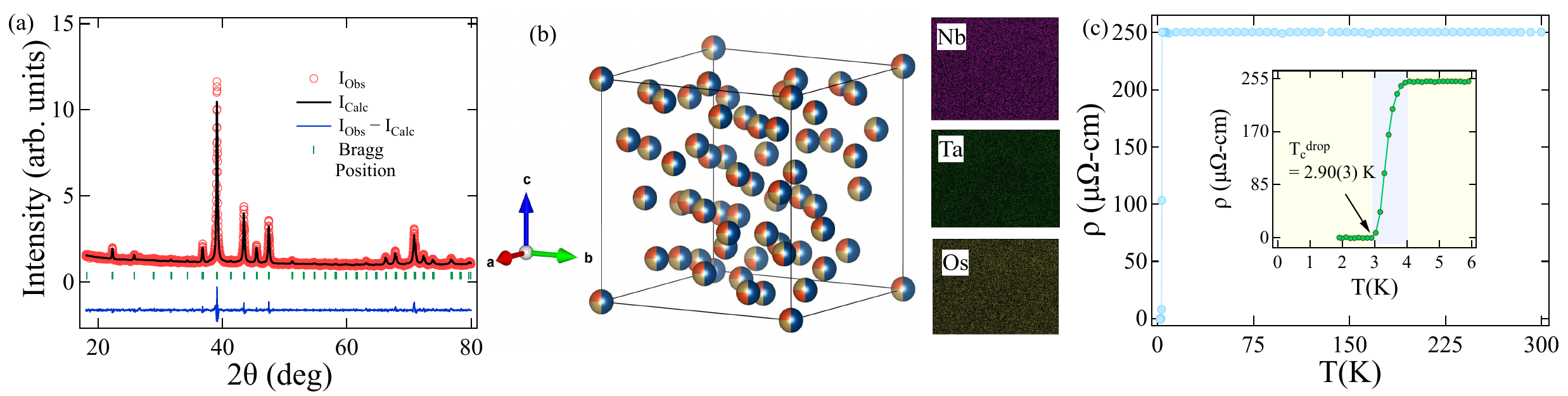}
\caption{\label{XRD} (a) Powder XRD pattern of NbTaOs$_{2}$ obtained at room temperature is indicated by the red circles; while the solid black lines represent the Rietveld refinement. (b) NbTaOs$_{2}$ crystal structure (left) and the elemental mapping (right) of Nb, Ta, and Os elements. (c) Temperature-dependent resistivity of NbTaOs$_{2}$, inset shows the superconducting transition temperature.}
\end{figure*}

\section{Experimental Details}
A polycrystalline sample of NbTaOs$_{2}$ was synthesized by melting high-purity metals (3N) Nb, Ta, and Os in a stoichiometric ratio in an argon atmosphere. Phase purity and crystal structure were analyzed using powder X-ray diffraction (XRD) with Cu~K$_{\alpha}$ radiation ($\lambda$=1.5406~\AA) on a PANalytical powder X-ray diffractometer and the XRD patterns were refined using Fullprof software~\cite{Fullprof}. Energy dispersive X-ray analysis (EDAX) was performed using an Oxford high-resolution field emission scanning electron microscope (SEM) to determine the elemental composition and obtain spatial elemental mapping. DC and AC susceptibility measurements were performed using a Quantum Design magnetic property measurement system (MPMS3). The electrical resistivity and heat capacity measurements were performed on a Quantum Design Physical Property Measurement System (PPMS). Muon spin rotation/relaxation ($\mu$SR) measurements were performed on the MUSR spectrometer at the ISIS pulsed neutron and muon facility at Rutherford Appleton Laboratory~\cite{Muonaidy}.

\section{Results}

\subsection{Sample Characterization}
The room-temperature powder XRD pattern at room temperature of NbTaOs$_{2}$ is presented in \figref{XRD}(a). Rietveld refinement confirms the phase purity, revealing that NbTaOs$_{2}$ crystallizes in the cubic structure $\alpha$-Mn with space group I$\Bar{4}3m$. The refined lattice parameters are a = b = c = 9.7703(2)~\text{\AA}, with a unit cell volume of V$_{cell}$ = 932.65(7)~\text{\AA}$^{3}$. \figref{XRD}(b) illustrates the NCS cubic crystal structure (left), where site mixing occurs at each lattice point. Furthermore, EDAX analysis verifies the elemental composition, showing Nb (24\%), Ta (24.5\%) and Os (51.5\%), which aligns closely with the nominal stoichiometric ratio of NbTaOs$_{2}$. Figure \ref{XRD}(b) represents the elemental mapping (right).

\begin{figure*} [t]
\includegraphics[width=2.0\columnwidth,origin=b]{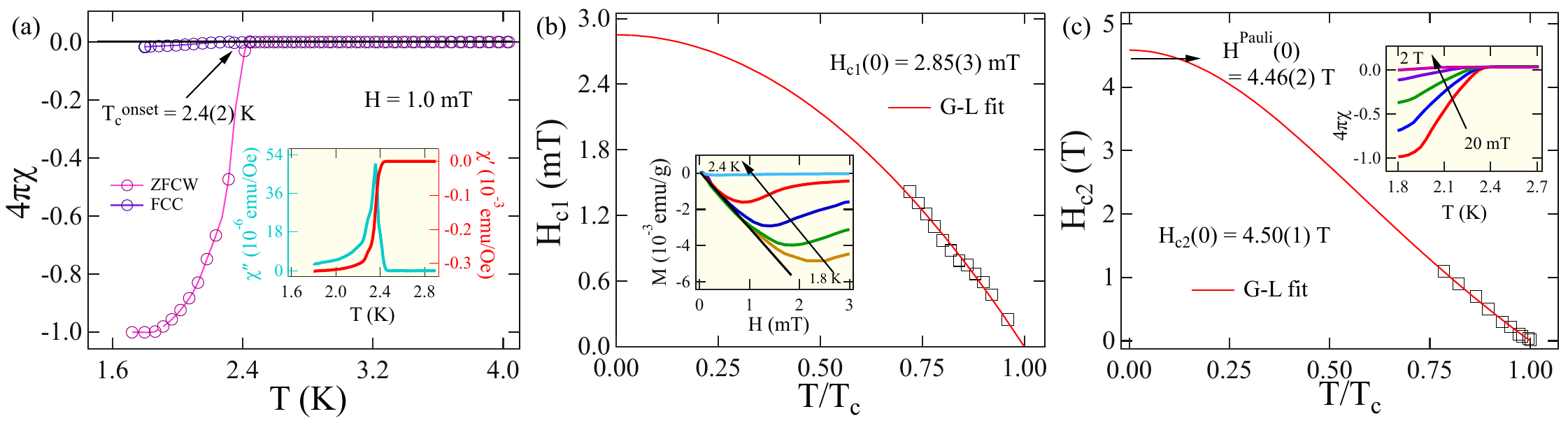}
\caption{\label{Mag}(a) Temperature $vs.$ magnetic susceptibility curves in ZFCW and FCC mode at 1~mT and the inset shows the real and imaginary parts of AC susceptibility. (b) Lower critical field $vs.$ reduced temperature (T/T$_c$) plot, the red curve represents the GL fit using \equref{Hc1} and the inset shows the field-dependent magnetization at different temperatures. (c) Upper critical field  $vs.$ T/T$_c$ plot, where the red curve represents GL fit using \equref{Hc2} and the inset shows the temperature $vs.$ susceptibility curves at different fields.}
\end{figure*}

\subsection{Resistivity}
Temperature-dependent electrical resistivity $\rho(T)$ measurements were performed, which confirm the superconducting transition in NbTaOs$_{2}$. The zero field $\rho(T)$ did not show a significant change in resistivity from room temperature (300~K) down to 4~K. The residual resistivity ratio (RRR) $\sim$ 1, suggests intrinsic disorder, which is also observed in other $\alpha$ -Mn binary and ternary compounds ~\cite{Re8NbTa}. The variation of $\rho(T)$ is shown in \figref{XRD} (c), while the inset highlights the emergence of zero resistivity at the transition temperature, $T_c$ = 2.90(3)~K. Compared to magnetization and specific heat measurements, the higher $T_c$ observed in resistivity measurements suggests the presence of surface superconductivity ~\cite{PhysRevB.88.144511}.

\subsection{Magnetization}
Temperature-dependent magnetization measurements were conducted on NbTaOs$_{2}$ sample using zero-field cooled warming (ZFCW) and field cooled cooling (FCC) modes [see \figref{Mag}(a)], confirming the bulk superconductivity with a transition temperature of $T_{c} = 2.4(2)$~K. Furthermore, measurement of AC susceptibility (inset of \figref{Mag}(a)) confirms $T_{c}$ of NbTaO$_{2}$.

Temperature and field-dependent magnetization measurements were performed to determine the superconducting parameters of NbTaOs$_{2}$. Using the Ginzburg-Landau (GL) relation (\equref{Hc1}), the lower critical field, $H_{c1}(0)$, was determined from the measured low field isothermal magnetization curves, where $H_{c1}$ at each temperature was identified as the point where the curves deviate from the Meissner line (solid black line) as shown in the inset of \figref{Mag}(b).
\begin{equation}
\label{Hc1}
H_{c1}(T) = H_{c1}(0)~[1-(T/T_{c})^{2}] 
\end{equation}
Using \equref{Hc1} $H_{c1}(0)$ was found to be 2.85(3) mT.

Further, the upper critical field $H_{c2}(0)$ was estimated by measuring the temperature-dependent susceptibility curves at different magnetic fields, and the onset of the diamagnetic signal is considered field-dependent $T_{c}$. $H_{c2}$ was plotted against temperature, which is well described by the GL relation (\equref{Hc2}), 
\begin{equation}
\label{Hc2}
H_{c2}(T) = H_{c2}(0)~\frac{(1-t^{2})}{(1+t^{2})};\quad  \text{where} ~t = T/T_{c}
\end{equation}
The GL fit using \equref{Hc2} is indicated by the red curve in \figref{Mag}(c), which yields an estimated value of $H_{c2}(0)$ = 4.5(1)~T.

Under an external magnetic field, Cooper pairs can break through two mechanisms: orbital pair breaking and the Pauli limiting field. The orbital limiting field is described by the Werthamer-Helfand-Hohenberg (WHH)~\cite{WHH1,WHH2} model, neglecting the effects of both spin-orbit scattering and Pauli paramagnetism, which is defined as the following equation
\begin{equation}
H_{c2}^{orb}(0)= -\alpha T_{c} \left.\frac{dH_{c2}(T)}{dT}\right\vert_{T=T_{c}}
\end{equation}
We estimate the initial slope of the upper critical field near the critical temperature, given by, $\left(\frac{-dH_{c2}(T)}{dT}\right)$, to be 0.89~T/K. In dirty limit superconductivity, the orbital limiting field can be approximated using the parameter $\alpha = 0.693$, yielding an estimated value of $H_{c2}^{orb}(0)$ = 1.5(1)~T. For conventional BCS superconductors, the Pauli limiting field is given by $H^{Pauli} = 1.86~T_{c}$~\cite{Pauli1,Pauli2}. The estimated $H^{Pauli}$ value is found to be 4.46(2)~T by using $T_{c}$ = 2.4(2)~K obtained from magnetization measurements. Interestingly, the estimated upper critical field, $H_{c2}(0)$ = 4.5(1)~T, is close to the Pauli pair-breaking limit. This suggests that spin-orbit scattering due to impurities or intrinsic spin-orbit interactions, predicted for non-centrosymmetric superconductors, may be influencing the system, leading to $H_{c2}(0)$/$H^{Pauli}$ $\ge$ 1~\cite{PhysRevB.75.184529,lu2015evidence,xi2016ising}. To gain insight into characteristic lengths, we employ Ginzburg-Landau (GL) theory to estimate the superconducting coherence length, $\xi_{GL}$(0)~\cite{tinkham} using the expression 
\begin{equation}
H_{c2}(0)=\frac{\Phi_{0}}{2\pi\xi_{GL}^{2}}
\end{equation}
where, $\Phi_{0}$ (= 2.07 $\times$10$^{-15}$ T m$^{2}$) is the magnetic flux quantum. Substituting $H_{c2}(0)$ = 4.5(1)~T, we obtain $\xi_{GL}$(0) = 85.4(4)~$\text{\AA}$.

The GL penetration depth, $\lambda_{GL}$(0)~\cite{lambda}, is estimated from the equation \equref{lambda}
\begin{equation}
H_{c1}(0)=\frac{\Phi_{0}}{4\pi\lambda_{GL}^2(0)}\left( ln \frac{\lambda_{GL}(0)}{\xi_{GL}(0)} + 0.12\right)
\label{lambda}
\end{equation}
Given $H_{c1}(0)$ = 2.85(3) mT, we determine $\lambda_{GL}$(0) = 4909(8)~$\text{\AA}$.
The resulting GL parameter defined as $k_{GL}$ = $\frac{\lambda_{GL}(0)}{\xi_{GL}(0)}$, is found to be $k_{GL}$ = 57.4(4), which is significantly greater than $\frac{1}{\sqrt{2}}$, classifying NbTaOs$_{2}$ as a strong type-II superconductor.
Furthermore, the thermodynamic critical field, $H_{c}(0)$, can be estimated using the relation $H_{c1}(0)H_{c2}(0) = H_{c}^2(0)ln(k_{GL})$. The estimated value of $H_{c}(0)$ was found to be 56.3(2)~mT for NbTaOs$_{2}$. The Maki parameter~\cite{Maki} is defined as the measure of the relative strength of the orbital and Pauli limiting effects in a superconductor. It is defined as $\alpha_{M} = \sqrt{2} \frac{H_{c2}^{orb}}{H_{c2}^{Pauli}}$. Substituting the calculated values, we obtain $\alpha_{M}$ = 0.47(2), indicating that the paramagnetic effect on pair breaking is negligible.
\begin{figure*} [t!]
\includegraphics[width=2.0\columnwidth,origin=b]{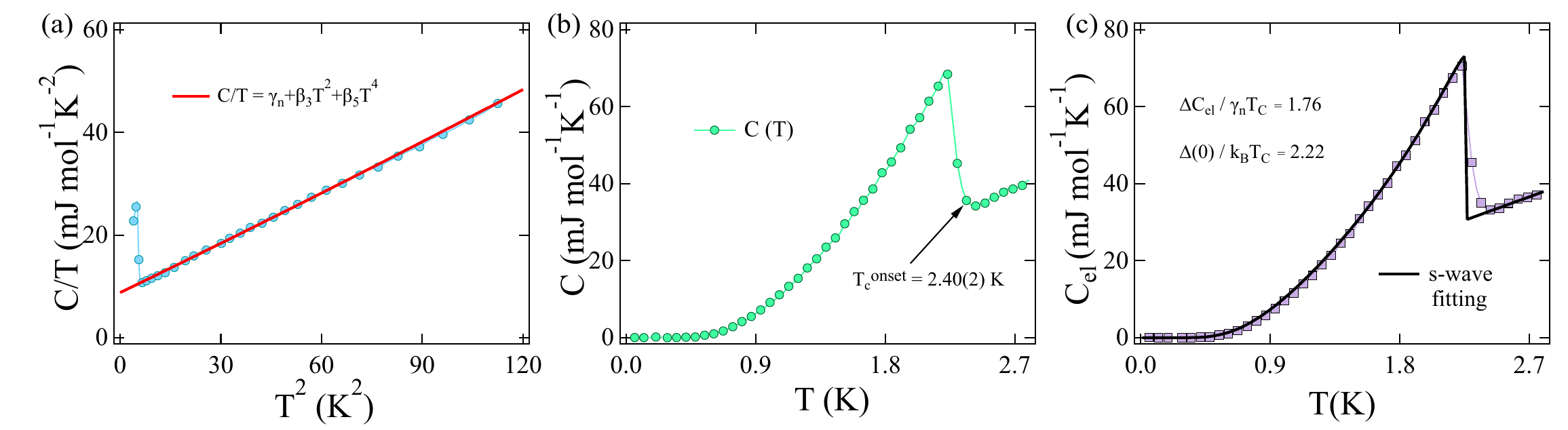}
\caption{\label{SH}(a) C/T $vs.$ T$^{2}$ plot, where the solid red line represents a fit to \equref{C/T} (b) Total specific heat $vs.$ temperature plot (c) Temperature-dependent electronic-specific heat data, fitted by the isotropic single gap model, and the solid black curve shows the s-wave fitting.}
\end{figure*}

It is essential to assess the stability of the vortex system against various factors that may disturb its equilibrium. The Ginzburg-Levanyuk number ($G_{i}$) quantifies the influence of thermal fluctuations on the process of vortex unpinning in a type-II superconductor~\cite{Gi} and is represented by \equref{Gi}.
\begin{equation}
\label{Gi}
G_{i}= \frac{1}{2} \left(\frac{k_{B}\mu_{0} \tau T_{c}}{4\pi \xi(0)^{3} H_{c}^{2}(0)}\right)^{2}
\end{equation}
Here, $\tau$ is the anisotropy factor (1 for cubic NbTaOs$_{2}$). We obtained $G_{i}$ = 1.45(4) $\times$ 10$^{-8}$ using the estimated parameters $T_{c}$ = 2.4(5)~K, $\xi(0)$ = 85.4(4)~\AA~ and $H_{c}(0)$ = 56.3(2)~mT. This value is close to low $T_{c}$ superconductors ($\sim$10$^{-8}$), suggesting that thermal fluctuations are not significantly responsible for the vortex unpinning in our compound~\cite{LowTcGi}.

\subsection{Specific Heat}
To further investigate the superconducting and normal state properties of NbTaOs$_2$, we performed a temperature-dependent specific heat $C(T)$ measurement. A distinct jump in specific heat was observed at $T_{c}$ = 2.40(2)~K, as shown in \figref{SH} (b), consistent with the onset of diamagnetic signal in magnetization measurements. The specific heat data plotted as $C/T$ vrs $T^2$ in \figref{SH}(a), analyzed using the Debye-Sommerfeld relation (\equref{C/T}) 
\begin{equation}
C/T=\gamma_{n}+\beta_{3}T^{2}+\beta_{5}T^{4}
\label{C/T}
\end{equation}

where $\gamma_{n}$ is the Sommerfeld coefficient (electronic contribution to specific heat), $\beta_{3}$ corresponds to the phononic contribution, and $\beta_{5}$ captures the higher-order anharmonic effects. The best fit of \equref{C/T} to the specific heat data in the normal state, shown as the red curve in \figref{SH}(a), gives the parameters: $\gamma_{n}$ = 8.78(1) mJ mol$^{-1}$K$^{-2}$, $\beta_{3}$ = 0.316(4) mJ mol$^{-1}$K$^{-4}$ and $\beta_{5}$ = 0.108(3) mJ mol$^{-1}$K$^{-6}$.

$\gamma_{n}$ is related to the density of states at the Fermi surface, $D(E_{F})$, through the following relation;
\begin{equation}
\gamma_{n} = \left(\frac{\pi^{2}k_{B}^{2}}{3}\right)D(E_{F}),
\end{equation}
where k$_{B}$ is the Boltzmann constant ($\approx$ 1.38 $\times$ 10$^{-23}$ J K$^{-1}$). Using this relation, the density of states at the Fermi level for NbTaOs$_{2}$ is estimated to be $D(E_{F}) = 3.73(3)$ states/eV f.u. 
Furthermore, the Debye temperature ($\theta_{Debye}$) can be estimated from the phononic coefficient $\beta_{3}$ using the Debye model,
\begin{equation}
\theta_{Debye} = \left(\frac{12\pi^{4}RN}{5\beta_{3}}\right)^{\frac{1}{3}},
\end{equation}
where $N$ represents the number of atoms per formula unit and $R = 8.314$ J mol$^{-1}$ K$^{-1}$ is the molar gas constant. Applying this formula, the Debye temperature for NbTaOs$_{2}$ is determined to be $\theta_{Debye} = 290(1)$~K.

The electron-phonon coupling strength can be evaluated using the dimensionless parameter $\lambda_{e-ph}$, derived from McMillan's model~\cite{McM}. This parameter is calculated using estimated values of $\theta_{Debye}$ and $T_{c}$ and is given by
\begin{equation}
\lambda_{e-ph} = \frac{1.04 + \mu^{}\ln(\theta_{Debye}/1.45T_{c})}{(1 - 0.62\mu^{})\ln(\theta_{Debye}/1.45T_{c}) - 1.04}
\end{equation}
Here, $\mu^{}$ represents the screened Coulomb repulsion, with a typical value of $\mu^{} = 0.13$ used for transition metals~\cite{McM}. Using $\theta_{Debye} = 290(1)$~K and $T_{c} = 2.25(3)$~K, the calculated electron-phonon coupling constant is $\lambda_{e-ph} = 0.53(4)$~\cite{sep1,sep2}.

Figure \ref{SH}(c) presents the temperature dependence of the electronic specific heat for NbTaOs$_{2}$. The electronic specific heat, $C_{el}$, is obtained by subtracting the phononic contribution, $C_{ph} = \beta_{3}T^{3} - \beta_{5}T^{5}$, from the total specific heat, $C$. The magnitude of the electronic specific heat jump in the superconducting transition, characterized by $\frac{\Delta C_{el}}{\gamma_{n}T_{c}}$, is found to be 1.76(3). This value exceeds the prediction value (1.43) of the BCS theory in the weak-coupling limit, indicating moderate electron-phonon coupling in NbTaOs$_{2}$.\\

\begin{figure*} [t!]
\includegraphics[width=2.0\columnwidth,origin=b]{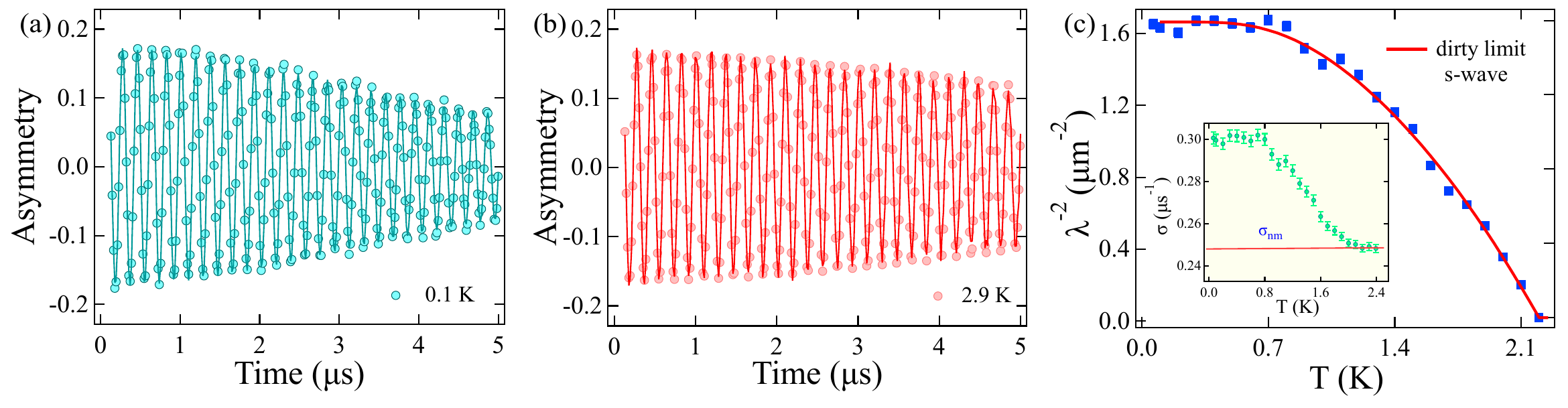}
\caption{\label{TF} TF-$\mu$SR spectra collected (a) below (0.1~K) and (b) above (2.9~K) $T_{c}$. (c) Temperature dependence of inverse square penetration depth ($\lambda^{-2}$), where the red solid curve shows the dirty-limit s-wave fitting.}
\end{figure*}

The symmetry of the superconducting gap can be examined by analyzing the temperature-dependent $C_{el}$. The relationship between normalized entropy ($S$) in the superconducting state and $C_{el}$ is expressed as:
\begin{equation}
C_{el} = t~\frac{dS}{dt}, \quad \text{where} ~t = T/T_{c}
\end{equation}

Within the BCS framework, the normalized entropy for a single isotropic gap is given by
\begin{equation}
\frac{S}{\gamma_{n}T_{c}} = -\frac{6}{\pi^2}\left(\frac{\Delta(0)}{k_{B}T_{c}}\right)\int_{0}^{\infty} \left[ f\ln(f)+(1-f)\ln(1-f) \right] dy,
\end{equation}
where $f = [\exp(E(\xi)/k_{B}T)+1]^{-1}$ is the Fermi function, and $E(\xi) = \sqrt{\xi^{2}+\Delta^{2}(t)}$. Here, $\xi$ represents the measured energy of normal electrons relative to the Fermi energy, and $y = \xi/\Delta(0)$.

The temperature dependence of the superconducting gap in the BCS approximation is described by
\begin{equation}
\Delta(t) = \tanh[1.82(1.018((1/t)-1))^{0.51}].
\end{equation} 

The solid black curve in \figref{SH}(c) represents the fit to the electronic specific heat data based on a single isotropic nodeless gap within the BCS model~\cite{BCS1,BCS2}. It provides $\Delta(0)/{k_{B} T_{c}} = 2.22(1)$, which is higher than the standard BCS value of 1.76, indicating moderate electron-phonon coupling in NbTaOs$_2$.

\subsection{Muon spin rotation and relaxation measurements}

\textbf{Transverse-field $\mu$SR}: We have performed $\mu$SR measurements in the transverse field (TF) geometry to investigate the nature of the superconducting gap. In the normal state, a magnetic field of 40~mT was applied perpendicular to the initial muon spin direction, ensuring that it was above H$_{c1}$ before cooling the sample to a base temperature of 0.1~K. The asymmetry spectra were recorded at different temperatures and the representative spectra measured below and above $T_{c}$ are shown in \figref{TF}(a) and (b), respectively. Furthermore, the time-domain spectra are accurately modeled using an oscillatory function combined with a Gaussian relaxation component as follows
\begin{equation}
 A_{\mathrm{TF}}(t) = \sum_{i=1}^{N} A_{i}~\mathrm{exp}\left(\frac{-\sigma^{2}_{i}t^{2}}{2}\right)\mathrm{cos}(w_{i}t+\phi)
\label{ATF}
\end{equation}
where $A_{i}$ is the initial asymmetry, $\sigma_{i}$ is the depolarization rate, $\phi$ is the initial phase and $\omega_i$ is the precession frequency. \equref{ATF} is used to fit the asymmetry spectra at different temperatures for $N = 2$, where $\sigma_{1}$ corresponds to muons interacting with the internal field distribution in the sample, while $\sigma_{2} = 0$ corresponds to the background (silver sample holder). 

The total depolarization rate ($\sigma$) is defined as
\begin{equation}
{\sigma}^2 = {\sigma_{fll}}^2 +{\sigma_{nm}}^2
\label{sigsc}
\end{equation}
where $\sigma_{fll}$ accounts for the contribution from FLL in the superconducting state, while $\sigma_{nm}$ corresponds to the relaxation due to randomly oriented nuclear dipole moments, which remains constant above $T_{c}$. The clear difference in $\sigma$ between the $T < T_{c}$ spectra and those recorded in the normal state (as illustrated in \figref{TF}(c)) is a direct consequence of the inhomogeneous field distribution induced by the FLL. In contrast, the weak Gaussian damping observed in the $T > T_{c}$ spectra arises from relaxation caused by randomly oriented nuclear dipolar fields~\cite{NbOs2,Re8NbTa}. The inset of \figref{TF}(c) shows the temperature dependence of $\sigma$ for NbTaOs$_{2}$ measured at 40~mT. The nuclear contribution to the depolarization rate, $\sigma_{nm} = 0.2454(3)~\mu s^{-1}$, is determined as the average depolarization rate above $T_{c}$. The superconducting contribution can then be extracted using \equref{sigsc}.
\begin{figure*} [t!]
\includegraphics[width=1.5\columnwidth,origin=b]{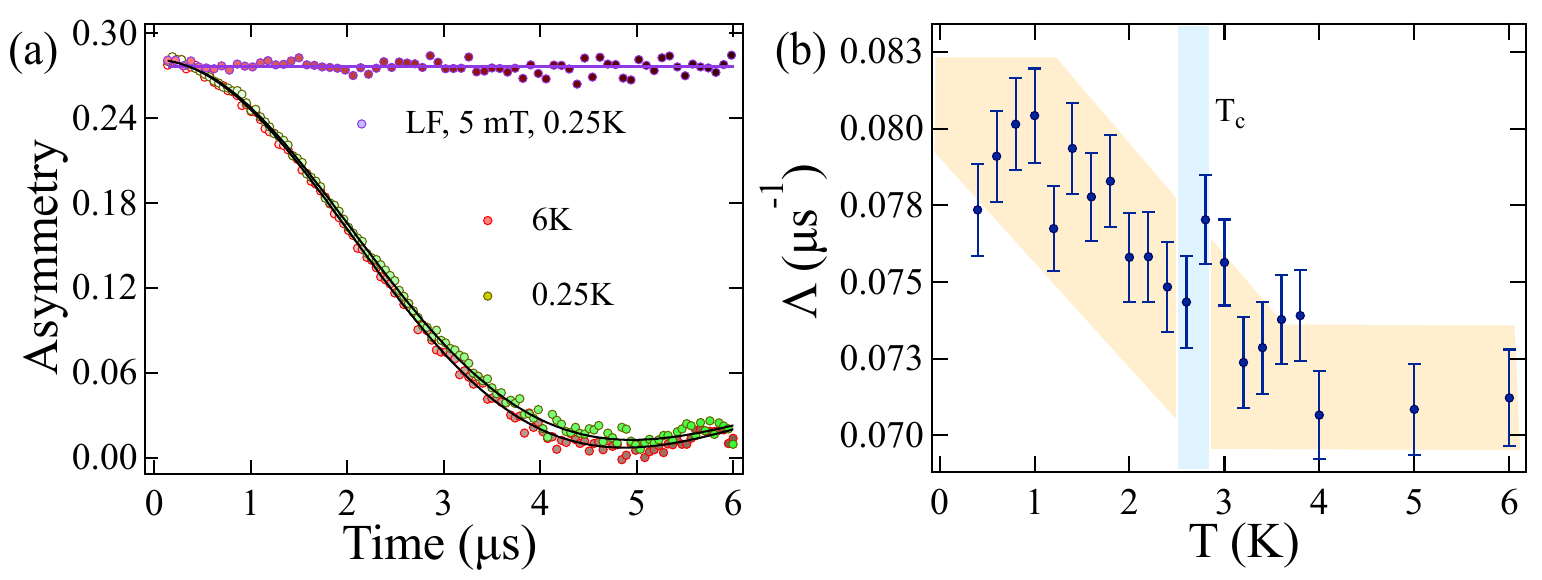}
\caption{\label{ZF}(a) ZF-$\mu$SR spectra collected below (0.25~K) and above (6~K) $T_{c}$ and fitted using the static Kubo-Toyabe function with additional function as in \equref{AZF}. (b) Temperature-dependence of additional relaxation rate $\Lambda$ with notable increase below $T_{c}$.}
\end{figure*}
Furthermore, a well-ordered triangular or hexagonal Abrikosov vortex lattice is expected to form in the mixed state. The temperature-dependent magnetic penetration depth $\lambda(T)$ can be constructed from $\sigma_{fll}(T)$, using the simplified Brandt formula~\cite{EHB} under the primary condition that $H/H_{c2} \gg 1$.
\begin{equation}
\label{lambdasc}
\sigma_{fll}(T) = \frac{0.0609 \times \gamma_{\mu} \phi_{0}}{\lambda^{2}(T)}
\end{equation}
where, $\phi_{0}$ is defined as the magnetic flux quantum and $\gamma_{\mu} = 2\pi \times$135.5~MHz/T is the muon gyromagnetic ratio. The temperature dependence of $\lambda^{-2}$, obtained from \equref{lambdasc}, is shown in \figref{TF}(c). To analyze the temperature-dependent behavior of $\lambda^{-2}$, we employ the following function within London's approximation for a dirty-limit BCS superconductor~\cite{Pene}
\begin{equation}
\frac{\sigma_{fll}^{-2}(T)}{\sigma_{fll}^{-2}(0)} = \frac{\lambda^{-2}(T)}{\lambda^{-2}(0)} = \frac{\Delta(T)}{\Delta(0)}~\mathrm{tanh}\left[\frac{\Delta(T)}{2k_{B}T}\right]
\label{London}
\end{equation}
The temperature dependence of the energy gap in the BCS approximation is given by
\begin{equation}
\frac{\Delta(T)}{\Delta(0)} = \tanh[1.82(1.018((\mathit{T_{c}/T})-1))^{0.51}]
\end{equation}
where $\Delta(0)$ represents the energy gap value at 0~K. 

The solid red curve in \figref{TF}(c) represents the fit obtained using \equref{London}, yielding $\Delta(0) = 0.38(2)$~meV at 40~mT. This results in a normalized superconducting gap value of $\Delta(0)/k_{B}T_{c} = 2.02(1)$, which closely matches the value determined from our specific heat measurements. Additionally, the magnetic penetration depth at 0~K is estimated as $\lambda(0) = 7321(2)$~$\text{\AA}$ at 40~mT using \equref{lambdasc}. This value is slightly higher than the $\lambda(0)$ estimated from the magnetization measurements; this difference arises from the different methods used to determine the depth of penetration~\cite{PhysRevB.98.214517}. TF-$\mu$SR and specific heat measurements indicate a fully open nodeless superconducting gap attributed to moderate electron-phonon coupling in NbTaOs$_{2}$.\\

\textbf{Zero-field $\mu$SR}: Zero-field (ZF) $\mu$SR measurements were performed to investigate the presence of spontaneous magnetic fields in the superconducting state of NbTaOs$_{2}$. Relaxation spectra were recorded in the absence of an external magnetic field. The ZF-$\mu$SR spectra collected above (T = 6~K) and below (T = 0.25~K) $T_c$ are presented in \figref{ZF}(a). To analyze the ZF-$\mu$SR data at various temperatures, we use the fitting function given in \equref{AZF}. This function consists of a static Gaussian-type Kubo-Toyabe (KT) relaxation term ($G_{KT}(t)$) combined with an additional relaxation component and a constant background. The solid black curves in \figref{ZF}(a) represent the fits to the data using \equref{AZF}.
\begin{equation}
\label{AZF}
A(t) = A_{i}~G_{KT}(t) e^{\left({-\Lambda t}\right)^{\beta}}+A_{bg}
\end{equation}

The KT function $G_{KT}(t)$, describes the asymmetry spectra associated with static and randomly oriented nuclear moments~\cite{SKT} and is given by
\begin{equation}
\label{KT}
G_{KT}(t)= \frac{1}{3}+\frac{2}{3} (1- {\Delta_{Z}^{2}t^{2}) \exp\left(\frac{-{\Delta_{Z}}^{2}~{t}^{2}}{2}\right)}
\end{equation}

Here, $\Delta_{Z}$ represents the relaxation rate due to nuclear dipolar fields, while $A_{i}$ denotes the initial asymmetry. The temperature-independent background asymmetry, $A_{bg}$, originates from muons stopping in the silver sample holder. $\Lambda$ corresponds to the additional relaxation rate, modified by a factor $\beta$. Across all temperature fits, $\beta$ remained approximately 2 during the fitting~\cite{fujihala2017possible}. As shown in \figref{ZF}(a), a subtle difference in the spectra above and below $T_c$ suggests enhanced relaxation of the muon spin polarization in the superconducting state. Notably, when the KT relaxation rate, $\Delta_{Z}$, $A_{i}$ and $A_{bg}$ remained fixed, the additional Gaussian relaxation rate, $\Lambda$, (for $\beta$ = 2) exhibits a significant increase below $T_c$ (see \figref{ZF}(b)), indicating the emergence of a weak spontaneous magnetic field. This relaxation rate remains nearly constant above $T_c$.

Further, longitudinal field (LF)-$\mu$SR measurements were conducted to rule out the possibility that the observed increase in relaxation originates from dilute impurities. Applying a 5~mT field parallel to the muon spin direction at 0.25~K effectively decoupled the muon spin polarization from internal magnetic fields, as indicated by the nearly flat spectra (purple) in \figref{ZF}(a). This confirms that the relaxation observed in ZF-$\mu$SR primarily arises from static internal fields. Additionally, the absence of oscillatory components in the LF spectra confirms the absence of magnetic ordering in our sample.

The magnitude of the internal magnetic field emerging below $T_c$ was estimated using the relation~\cite{Re6Ti}
\begin{equation} 
B_{int} = \frac{\delta\Lambda}{\gamma_{\mu}} 
\end{equation}
where $\delta\Lambda$ represents the increase in the additional relaxation rate ($\Lambda$) observed below $T_c$, with parameters $\Delta_{Z}$ and $\beta$ held constant. This increase was determined to be 0.008(1)~$\mu$s$^{-1}$, yielding an internal field of $B_{int}$ = 0.094(5)~G, and it is consistent with the values reported for NCS exhibiting TRSB~\cite{Re6Ti,La7Rh3}.

Our observation establishes NbTaOs$_{2}$ as the first member of the Re-free $\alpha$-Mn cubic non-centrosymmetric (NC) structural family to exhibit time-reversal symmetry breaking (TRSB), given that previous $\mu$SR investigations have not detected TRSB in other Re-free $\alpha$-Mn NC superconductors~\cite{NbOs,NbOs2,TaOs,Mg10Ir19B16}. This finding challenges the assumption that Re is crucial for inducing TRSB in Re-based $\alpha$-Mn superconductors like Re$_6$X (where X = Ti, Zr, Hf), where TRSB is commonly observed. The weak spontaneous magnetic field in NbTaOs$_2$ could be due to enhanced antisymmetric spin-orbit coupling (ASOC), possibly arising from the mixed 4d/5d sites within the material. Although mixed sites can introduce disorder, a comparable residual resistivity ratio (RRR) in NbTaOs$_2$ to other NC binary superconductors suggests that TRSB in NbTaOs$_{2}$ is not driven by disorder, as recently proposed by Andersen et al.~\cite{disorderTRSB}. However, a complete understanding of the roles of SOC, site mixing, and intrinsic disorder in the $\alpha$-Mn crystal structure requires further microscopic characterization of NbTaOs$_2$ samples with varying degrees of disorder, along with detailed electronic structure calculations.

\begin{figure}
\includegraphics[width=0.95\columnwidth,origin=b]{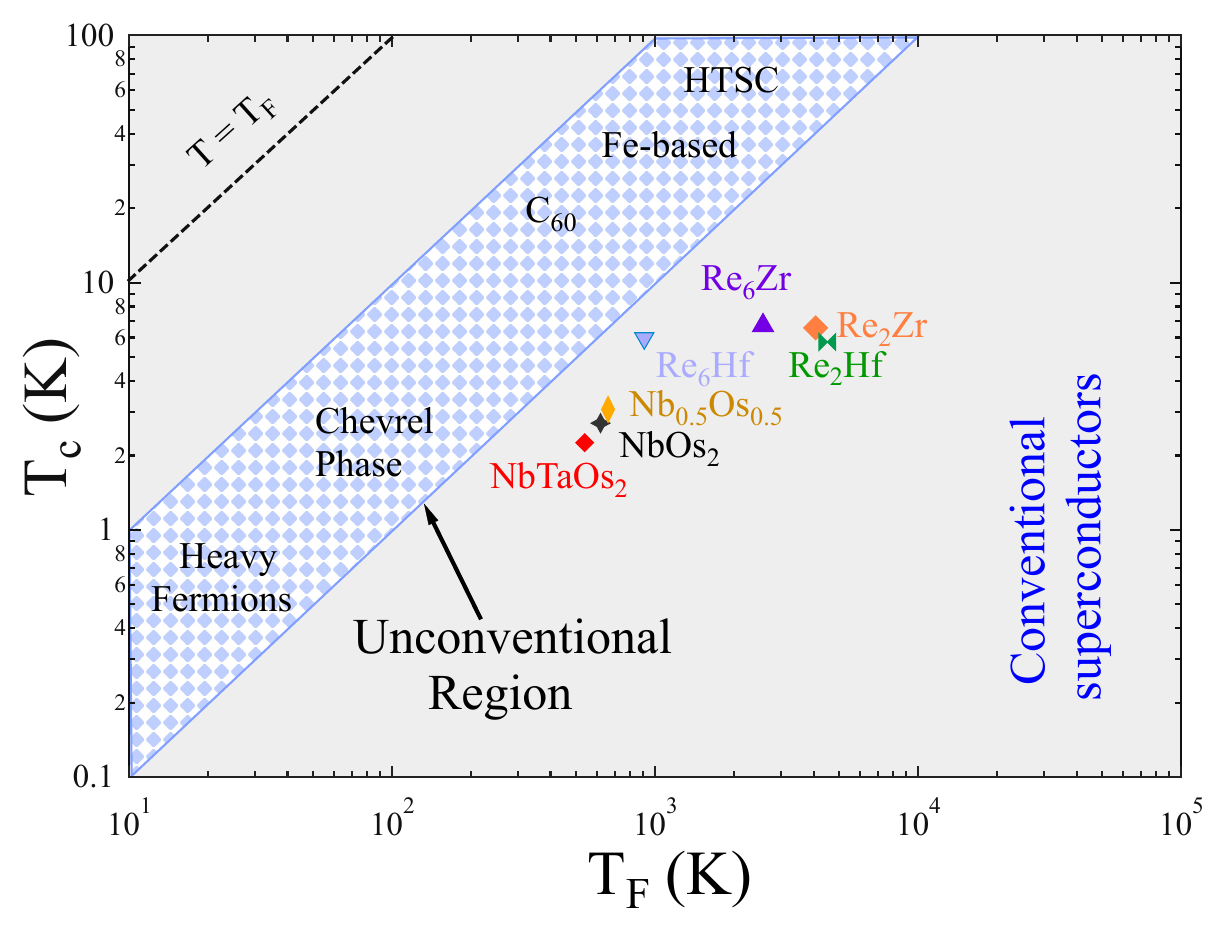}
\caption{\label{Uemura}T$_c$ $vs.$ T$_F$ plot for Uemura classification where the red diamond represents NbTaOs$_{2}$.}
\end{figure}

\section{Uemura Plot}
Uemura et al.~\cite{U1} proposed a classification for superconductors based on the ratio of their superconducting transition temperature ($T_c$) to the Fermi temperature ($T_F$). Unconventional superconductors have ratios between $0.01 \leq T_c/T_F  \leq 0.1$, distinguishing them from conventional superconductors, which have a significantly lower ratio. The Fermi temperature can be estimated using the expression $k_B T_F = \frac{\hbar^2}{2m_e} \left( 3\pi^2 n_s \right)^{2/3} \frac{1}{\left(1 + \lambda_{e\text{-}ph}\right)}$,
where $k_B$ is the Boltzmann constant, $\hbar$ is the reduced Planck constant, $m_e$ is the electron mass, $n_s$ is the density of super-electrons, and $\lambda_{e\text{-}ph}$ represents the electron-phonon coupling constant.
\begin{table} [t!]
\caption{Superconducting and normal state parameters of NbTaOs$_{2}$}
\label{tb2}
\begin{tabular}[b]{l c c }\hline\hline
Parameters& Unit &NbTaOs$_2$\\
\hline
$T_{c}$& K&2.4(2)\\
$H_{c1}(0)$&mT &2.85(3)\\
$H_{c2}(0)$& T&4.5(1)\\
$H_{c2}(0)^{P}$& T& 4.46(2)\\
$H_{c2}(0)^{orb}$(T)& T &1.5(1)\\
$\Delta C_{el}/\gamma_{n}T_{C}$& &1.76(3)\\
$\Delta(0)/k_{B}T_{C}$& &2.22(1)\\
$\gamma_{n}$&mJ mol$^{-1}$K$^{-2}$ &8.78(1)\\
$\theta_{D}$& K& 290(1)\\
$\lambda_{e-ph}$& &0.53(4)\\
$\xi_{GL}(0)$& $\text{\AA}$ &85.4(4)\\
$\lambda_{GL}(0)$& $\text{\AA}$ & 4909(8)\\
$k_{GL}$& &57.4(4)\\
$\lambda^{muon}(0)$& $\text{\AA}$ &7321(2)\\
$n_s$& $10^{26}$ $m^{-3}$& 0.74(2)\\
$T_{F}$& K& 539.5(2)\\
\hline\hline
\end{tabular}
\par\medskip\footnotesize
\end{table}
The density of super-electrons is related to the magnetic penetration depth ($\lambda$) and is given by~\cite{NbOs2}
\begin{equation}
n_s = \frac{m_e (1+\lambda_{e-ph})}{\mu_0 e^2 \lambda^2}
\label{ns}
\end{equation}
Using $\lambda_{e-ph}$ estimated from specific heat and $\lambda$ calculated from muon analysis in \equref{ns} yields $n_s = 0.74(2) \times 10^{26}$ m$^{-3}$.

Using the evaluated value of $n_s$ from \equref{ns} in the above expression for $T_F$, yielding $T_F$ = 539.5(2)~K, further provides the ratio $T_c/T_F$ = 0.0044, which places NbTaO$_2$ just outside the unconventional region, as shown by the red diamond in \figref{Uemura}. All superconducting and normal state parameters of NbTaOs$_2$ obtained from different techniques are listed in \tableref{tb2}.

\section{Conclusion}
In summary, we have successfully synthesized the Re-free ternary NbTaOs$_{2}$ with an $\alpha$-Mn structure. Our comprehensive measurements, such as electrical resistivity, magnetization, and specific heat, confirm that it exhibits type-II bulk superconductivity. Electronic specific heat and transverse field $\mu$SR data point to a moderately coupled s-wave superconducting gap structure. Intriguingly, the zero-field $\mu$SR measurements indicate time-reversal symmetry breaking (TRSB) in the superconducting ground state of NbTaOs$_{2}$. This marks the first observation of TRSB in a Re-free $\alpha$-Mn compound, which we believe is due to the enhanced spin-orbit coupling resulting from 4d/5d side mixing. This study not only paves the way for inducing TRSB in superconducting ground states but also offers further insights into the role of disorder in superconductivity, a topic that remains poorly understood.

\section{Acknowledgement}
R.K.K. thanks the UGC, Government of India, for a Senior Research Fellowship. R.P.S. acknowledges the SERB Government of India (Core Research Grant CRG/2023/000817). We also thank ISIS, STFC, UK, for $\mu$SR beamtime (RB: 2410330).

\bibliography{references}

\end{document}